\documentclass[iop]{emulateapj}
\usepackage{amsmath, amssymb}
\usepackage{amsfonts}
\usepackage{epsfig}
\usepackage[figuresright]{rotating}

\begin{document}

\def\simeq{
\mathrel{\raise.3ex\hbox{$\sim$}\mkern-14mu\lower0.4ex\hbox{$-$}}
}

%% our macros
\def\ltsima{$\; \buildrel < \over \sim \;$}
\def\simlt{\lower.5ex\hbox{\ltsima}}
\def\gtsima{$\; \buildrel > \over \sim \;$}
\def\simgt{\lower.5ex\hbox{\gtsima}}

\def\lsun{{\rm L_{\odot}}}
\def\msun{{\rm M_{\odot}}}
\def\rsun{{\rm R_{\odot}}} 
\def\lta{\la}
\def\gta{\ga}
\def\be{\begin{equation}}
\def\ee{\end{equation}}
\def\le{{L_{\rm Edd}}}
\def\rp{{R_{\rm ph}}}
\def\rs{{R_{\rm s}}}
\def\mo{{\dot M_{\rm out}}}
\def\me{{\dot M_{\rm Edd}}}
\def\tc{{t_{\rm C}}}
\def\rc{{R_{\rm core}}}
\def\mc{{M_{\rm core}}}
\def\mbh{{M_{\rm BH}}}
\def\e{{\dot m_{\rm E}}}
\def\del#1{{}}

\title{BAL QSOs and Extreme UFOs: the Eddington connection}
\author{Kastytis~Zubovas\altaffilmark{1}$^,$\altaffilmark{2} and Andrew~King\altaffilmark{1}}

\altaffiltext{1} {Theoretical Astrophysics Group, University of
Leicester, Leicester LE1 7RH, U.K.}
\altaffiltext{2} {Center for Physical Sciences and Technology, Savanori\c{u} 231, Vilnius LT-02300, Lithuania; kastytis.zubovas@ftmc.lt}

\begin{abstract}

We suggest a common physical origin connecting the fast, highly
ionized winds (UFOs) seen in nearby AGN, and the slower and less
ionized winds of BAL QSOs. The primary difference is the mass loss
rate in the wind, which is ultimately determined by the rate at which
mass is fed towards the central supermassive black hole (SMBH) on
large scales. This is below the Eddington accretion rate in most UFOs,
and slightly super--Eddington in extreme UFOs such as PG1211+143, but
ranges up to $\sim 10-50$ times this in BAL QSOs.  For UFOs this
implies black hole accretion rates and wind mass loss rates which are
at most comparable to Eddington, giving fast, highly--ionized winds.
In contrast BAL QSO black holes have mildly super--Eddington accretion
rates, and drive winds whose mass loss rates are significantly
super--Eddington, and so are slower and less ionized. This picture
correctly predicts the velocities and ionization states of the
observed winds, including the recently--discovered one in SDSS
J1106+1939. We suggest that luminous AGN may evolve through a sequence
from BAL QSO through LoBAL to UFO--producing Seyfert or quasar as
their Eddington factors drop during the decay of a bright accretion
event. LoBALs correspond to a short--lived stage in which the AGN
radiation pressure largely evacuates the ionization cone, but before
the large--scale accretion rate has dropped to the Eddington value.
We show that sub--Eddington wind rates would produce an $M - \sigma$
relation lying above that observed. We conclude that significant SMBH
mass growth must occur in super--Eddington phases, either as BAL QSOs,
extreme UFOs, or obscured from direct observation.
\end{abstract}


\section{Introduction}

Since their discovery more than a decade ago interest in the scaling
relations between supermassive black holes (SMBH) and their galactic
hosts has grown sharply. A promising candidate for the agency linking
the growth of the hole with the evolution of the host is a powerful
quasi-spherical wind from the central regions of an active galactic
nucleus. Examples of these are widely observed. The fastest,
(sometimes called UFOs) are seen in a large fraction of local AGN
\citep[][]{Tombesi2010A&A, Tombesi2010ApJ}. They are detected via
blueshifted X--ray absorption lines (typically heliumlike iron) with
velocities $\sim 0.03-0.15 c$ and very high ionization parameters $\xi
\gtrsim 10^4$ \citep[][]{Pounds2003MNRASb, Pounds2003MNRASa,
  Tombesi2010A&A, Tombesi2010ApJ, Chartas2002ApJ,
  Chartas2003ApJ}. Broad absorption line (BAL) QSOs
\citep[][]{Hazard1984ApJ, Knigge2008MNRAS, Gibson2009ApJ} form a
family with lower velocities and lower ionization. They comprise $\sim
10-20 \%$ of all QSOs \citep{Hazard1984ApJ, Knigge2008MNRAS,
  Gibson2009ApJ} and exhibit broad blueshifted absorption lines, with
low--velocity edge $v_{\rm min} \simeq 0.007c - 0.03c$ and
high--velocity edge $v_{\rm max} \simeq 0.01c - 0.06c$
\citep{Gibson2009ApJ}. The typical ionization states in BAL QSO
spectra are CIV, OVI, NV, with a small fraction, called LoBALs, having
low-ionization lines of MgII and FeII.

It is clear that at least some observed winds have the properties
needed to make the host galaxy sensitive to the growth of its central
black hole. The quasar PG1211+143 is a UFO whose observed velocity and
ionization parameter imply a momentum rate (thrust)
\begin{equation}\label{windmom}
\dot{M}_{\rm w} v_{\rm w} \simeq \frac{\le}{c},
\end{equation}
where $\dot{M}_{\rm w}, v_{\rm w}$ are the mass outflow rate and
velocity of the wind, and $\le$ is the Eddington luminosity
\citep[][]{Pounds2003MNRASb, King2003MNRASb}. This relation suggests
that the wind here is driven by photons of the AGN radiation field
Thomson scattering once before escaping the system \citep{King2003MNRASb,
King2010MNRASa}, A wind with this property must have a major effect
on the host galaxy, as it inevitably shocks against its interstellar
gas.

This offers an obvious way for the huge binding energy of a
supermassive black hole (SMBH) to affect its host, and so lead to an
explanation of the $M - \sigma$ relation. For black hole
masses $M$ below a critical value \citep{King2003ApJ,King2005ApJ}
\begin{equation}\label{eq:msigma}
M_{\sigma} = {f_g\kappa\over \pi G^2}\sigma^4 \simeq 3.7 \times 10^8
\sigma^4_{200} \; \msun
\end{equation}
(where $f_g$ is the gas fraction, $\kappa$ the electron scattering
opacity, and $\sigma$ the velocity dispersion of the host spheroid)
the Eddington thrust of the black hole wind is too weak to lift the
host interstellar gas far from the hole. The wind shock is efficiently
Compton--cooled by the black hole's radiation field and falls back
after sweeping up only small mass of interstellar gas. In constrast,
once the hole grows to a mass $M > M_{\sigma}$ the Eddington thrust
drives this shock far enough from the hole that it no longer cools,
expanding adiabatically instead. This powerful energy injection
efficiently sweeps up and expels most of the host's interstellar gas. This is
the probable cause \citep{Zubovas2012ApJ} of the observed high--speed
galaxy--wide molecular outflows \citep{Feruglio2010A&A, Rupke2011ApJ,
  Sturm2011ApJ} which clear galaxy spheroids of gas and make them red
and dead.

In this paper we quantitatively explore the possible range of
velocities and ionization equilibria in AGN winds. Our results suggest
that BAL QSO winds and extreme UFOs like PG1211+143 are manifestations
of the same physical process. The primary difference between the two
classes is the mass outflow rate in the wind, which has its ultimate
origin in the mass {\it inflow} rate towards the accreting
supermassive black hole from large scales. In BAL QSOs, this
large-scale inflow rate is highly super-Eddington, leading to a dense
and slow wind outflow, while extreme UFOs have more moderate inflow
rates $\dot{M}_{\rm acc} \sim \dot{M}_{\rm Edd}$. Variations in the
geometry of the accretion flow can account for the large spread in
outflow velocities for the same ionization species or the same object,
and vice versa. All these winds transmit momentum rates $\ga \le/c$ to
the host ISM, and so would correctly predict the observed $M - \sigma$
relation.

In contrast, the majority of UFOs have mass inflow rates which are
significantly sub--Eddington.  If these rates prevailed throughout the
growth of the SMBH they would produce either an $M - \sigma$ relation
lying significantly above what is seen, or no relation at all. We
conclude that significant SMBH mass growth occurs in (super) Eddington
phases. These manifest themselves either as BAL QSOs or extreme UFOs,
or correspond to obscured systems.

This paper is structured as follows. We review the properties of AGN winds in
Section \ref{sec:winds}. We then describe the model for calculating the
relation between wind ionization and velocity, and present the
results (Section \ref{sec:model}). Finally, we discuss the physical
interpretation and implications of our findings in Section \ref{sec:discuss}.

\section{AGN winds} \label{sec:winds}

The launching of winds from accreting AGN is discussed in
\citet{King2010MNRASa}. Here we present a brief overview,
concentrating on the properties relevant for the connection between
UFOs and BAL QSOs: wind ionization, velocity and mass outflow rate. We
first consider the likely range of accretion rates.

\subsection{Accretion rates in AGN}

In an accretion flow around a supermassive black hole (SMBH), the
accretion rate at each radius is determined by local properties and
the ultimate process feeding it, rather than directly by the black
hole itself. There is no reason that this rate should respect the
Eddington limit for the black hole -- at radii larger than the point
of direct infall to the hole, the Eddington limit may be significantly
exceeded. The properties of the host galaxy set a limit on the infall
rate $\dot M_{\rm in}$ on to the disk at large radii. This cannot be
larger than the rate given by allowing gas previously in equilibrium
suddenly to fall freely: for a roughly isothermal equilibrium with
velocity dispersion $\sigma \equiv 200 \sigma_{200}$~km/s this
dynamical rate is
\begin{equation}
\dot{M}_{\rm dyn} \sim \frac{f_{\rm g}\sigma^3}{G} \simeq 2 \times 10^3 \;
\sigma_{\rm 200}^3 \;\msun\; {\rm yr^{-1}} 
\end{equation}
where $f_{\rm g}$ is the gas fraction, i.e. the ratio of gas density in
  the galaxy to the total density; $f_{\rm g} \simeq 0.16$ is the cosmological
  value. Since $\dot M_{\rm in} < \dot M_{\rm dyn}$ this shows that
\begin{equation}\label{edd}
\frac{\dot M_{\rm in}}{\dot M_{\rm Edd}} < 40\frac{M_\sigma}{M
\sigma_{200}},
\end{equation}
where $\dot{M}_{\rm Edd} \simeq 2M_8\msun\,{\rm yr}^{-1} $ is the
Eddington accretion rate (with $M_8 = M/10^8\msun$ and accretion
efficiency $\eta \simeq 0.1$), $M_\sigma \simeq 3.67 \times 10^8
\sigma_{200}^4 \; \msun$ is the critical black hole mass
\citep{King2010MNRASa,Zubovas2012MNRASb} and M is the current black
hole mass. Clearly unless the black hole mass is significantly below
$M_\sigma$, or the galaxy has $\sigma \ll 200~{\rm km\,s}^{-1}$, the
maximum mass inflow rate feeding the accretion disk is $\lesssim 50$
times the Eddington limit. Since the dynamical rate is an extreme
upper limit this suggests that AGN are never fed at very high
Eddington rates.

\subsection{Wind launching in super-Eddington systems}

An accretion disk reacts to a locally super--Eddington inflow by
driving away the excess as a wind at each
radius. \citet{Shakura1973A&A} show that this mass loss implies that
the total luminosity of the system is
\begin{equation} \label{eq:ldisc}
L = L_{\rm Edd} \left(1 + {\rm ln}\left(1 + \dot{m}\right)\right),
\end{equation}
where
\begin{equation}
\dot{m} \equiv \frac{\dot{M}_{\rm w}}{\dot{M}_{\rm acc}} = \frac{\dot
  M_{\rm in}}{\dot M_{\rm acc}} - 1,
\end{equation}
with $\dot M_{\rm w}$ the wind outflow rate and $\dot M_{\rm acc}$ the black
hole accretion rate.

We have seen above that the accretion rate and the wind are never very
super--Eddington, so the wind optical depth to electron scattering is
$\sim 1$. This means that each photon emitted by the accretion on to
the black hole scatters about once before escaping. As electron
scattering is front--back symmetric, this shows that the wind acquires
a momentum flow rate
\begin{equation} \label{eq:eddthrust}
\dot{M}_{\rm w} v_{\rm w} \simeq \frac{L}{c},
\end{equation}
\citep{King2010MNRASa}
where $L$ is the total black hole and disk luminosity
(eq. \ref{eq:ldisc}).

\subsection{Wind launching in sub--Eddington systems}

In a sub-Eddington AGN, Compton scattering is by definition incapable
of launching winds. However, radiation may still be able to expel
matter in a wind by line--driving (see below): hot stars achieve this,
for example. For disk accretion this process is formally identical to
the super--Eddington case considered in the last subsection, except
that the Eddington luminosity is replaced by a lower luminosity
$L_{\rm crit}$. The analogous equation to (\ref{eq:ldisc}) holds, i.e.
\begin{equation} \label{eq:ldisc2}
L = L_{\rm crit} \left(1 + {\rm ln}\left(1 + \dot{m}\right)\right),
\end{equation}
where again
\begin{equation}
\dot{m} \equiv \frac{\dot{M}_{\rm w}}{\dot{M}_{\rm acc}} = \frac{\dot
  M_{\rm in}}{\dot M_{\rm acc}} - 1,
\end{equation}
The analogy with the super--Eddington case goes deeper.
Line--driving, i.e.  scattering off bound electrons, has to accelerate
the wind to the escape velocity from the photosphere in the immediate
vicinity of the accreting black hole. Thus absorption lines whose rest
energies lie below the peak of the continuum spectrum (largely a black
body at the effective temperature of about $10^5$\; K) are blueshifted
across most of the spectrum and so absorb and re--radiate
(i.e. scatter) almost all of the bolometric luminosity. In a similar
manner to equation (\ref{eq:eddthrust}) this gives
\begin{equation}\label{eq:subeddthrust}
\dot{M}_{\rm w} v_{\rm w} \simeq \frac{L_{\rm bol}}{c},
\end{equation}
which is essentially the argument of \citet{Cassinelli1973ApJ}. Figure
4 of \citet{Tombesi2013MNRAS} suggests that this approximate relation
may hold for a significant number of UFOs.

\subsection{Wind properties}

\begin{figure}
  \centering
    \includegraphics[width=0.45 \textwidth]{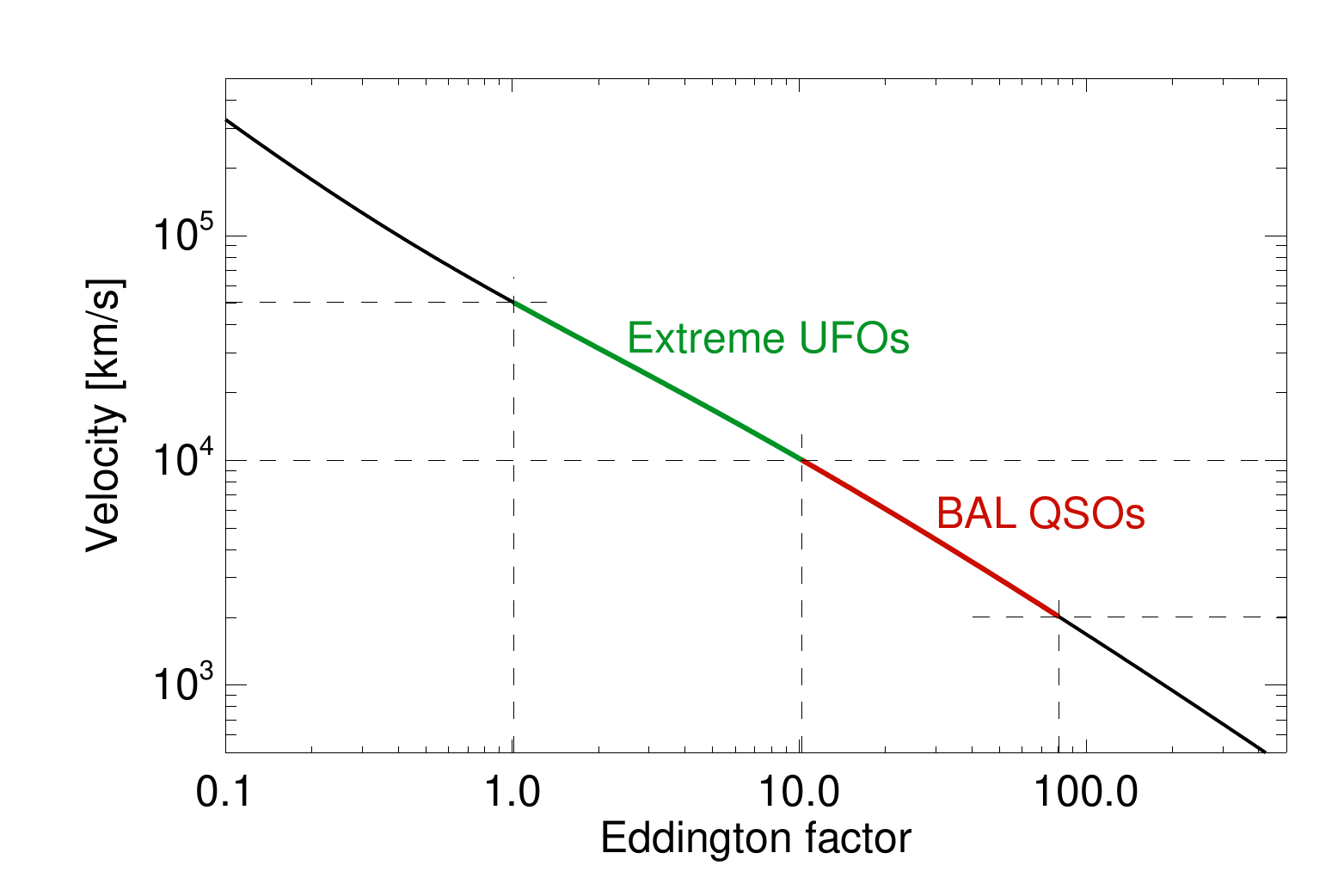}
    \includegraphics[width=0.45 \textwidth]{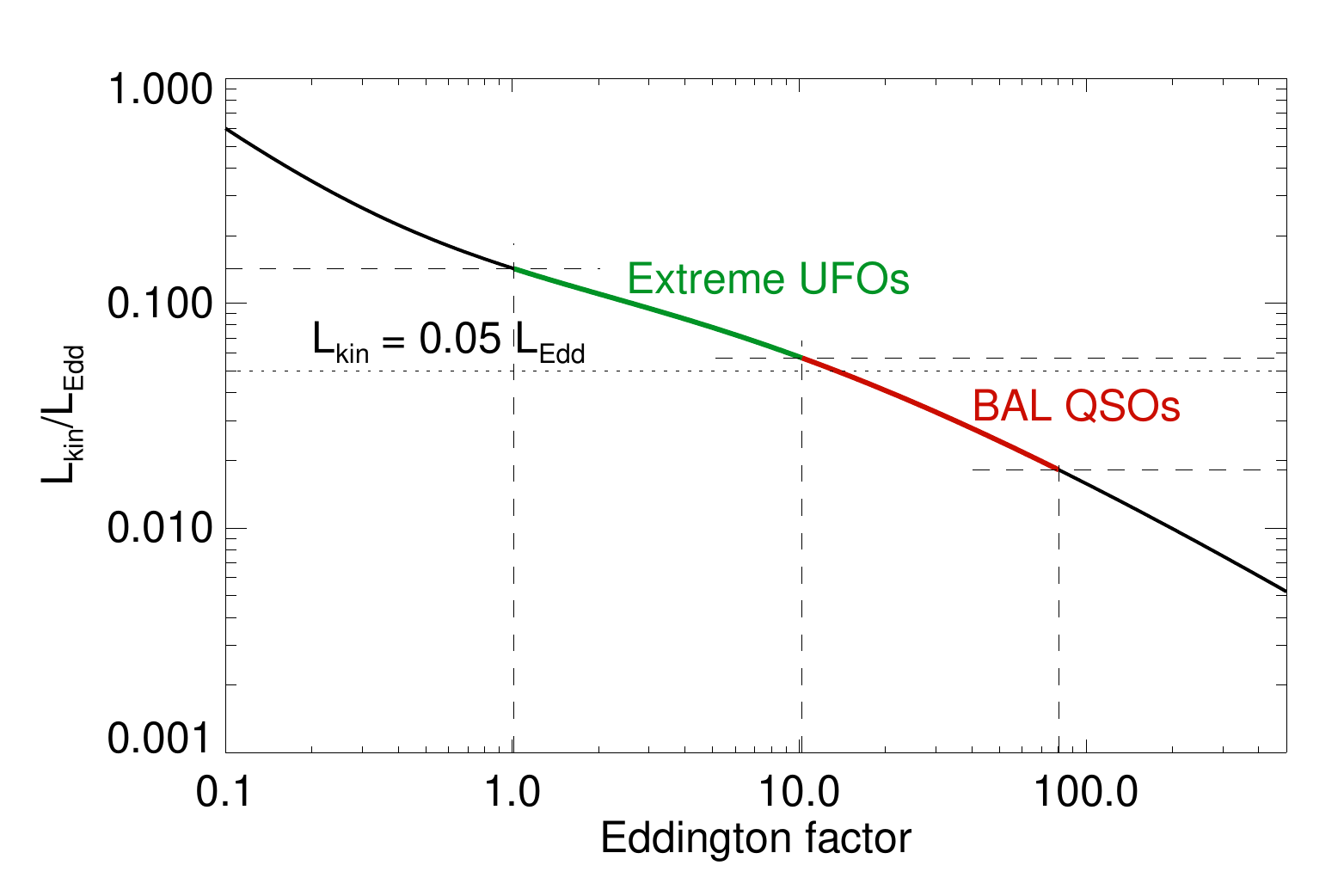}
  \caption{{\em Top}: Wind velocity as function of the Eddington factor, from
    eq. (\ref{eq:vwind}). Horizontal dashed lines show approximate limits of
    observed velocities of UFOs ($10^4 - 5\times10^4$~${\rm
      km\,s}^{-1}$) and BAL QSO outflows ($2000-10000$~${\rm km\,s}^{-1}$),
    with corresponding $\dot m$ factors identified. {\em Bottom panel}: Wind
    kinetic luminosity as function of the Eddington factor, from
    eq. (\ref{eq:lkin}).}
  \label{fig:mdotvel}
\end{figure}

The wind velocity $v_{\rm w}$ in both sub- and super-Eddington systems can be
expressed in terms of $\dot{m}$:
\begin{equation} \label{eq:vwind}
v_{\rm w} = \frac{\eta \dot{M}_{\rm acc} c^2[1 + {\rm ln}\left(1 +
  \dot{m}\right)]}{\dot{M}_{\rm w} c} = \eta c \frac{1 + {\rm
    ln}\left(1 + \dot{m}\right)}{\dot{m}}.
\end{equation}
The wind kinetic luminosity is
\begin{equation} \label{eq:lkin}
L_{\rm kin} = \frac{1}{2} \dot{M}_{\rm w} v_{\rm w}^2 = \frac{\eta}{2} L
\frac{[1 + {\rm ln}\left(1 + \dot{m}\right)]^2}{\dot{m}}.
\end{equation}
We plot these two relations in Figure \ref{fig:mdotvel} for $\eta =
0.1$ and $L = L_{\rm Edd}$, declining below this value for UFOs.  We
identify the typical ranges of UFO and BAL QSO outflow velocities
(green and red respectively) together with the corresponding $\dot m$
factors and kinetic luminosities; we note, however, that there is an
overlap in velocity between the two populations, as described in the
Introduction.

The gas density in the outflowing wind follows from the mass
outflow rate at radius $R$:
\begin{equation}
n_{\rm w} = \frac{\dot{M}_{\rm w}}{4 \pi R^2 m_{\rm p} v_{\rm w}}.
\end{equation}
with $m_{\rm p}$ the proton mass. UFOs are generally believed
to have large covering factors \citep[i.e. $\Omega_{\rm w}/2\pi
  \ga 0.6$,][]{Tombesi2010A&A,Tombesi2010ApJ} so we do not consider
potential outflow collimation. This gives the wind ionization
parameter far from the hole as:
\begin{equation} \label{eq:xi}
\begin{split}
\xi_{\rm w} = & \frac{L_{\rm ion}}{n_{\rm w}R^2} = \frac{b l_{\rm i} L \times
  4 \pi m_{\rm p} v_{\rm w}}{\dot{M}_{\rm w}} \\ & \simeq 4 \pi m_{\rm p}
\eta^2 c^3 \frac{1 + {\rm ln}\left(1 + \dot{m}\right)}{\dot{m}^{2}} b l_{\rm
  i},
\end{split}
\end{equation}
where $l_{\rm i}$ is the fraction of the AGN luminosity capable of ionizing a
particular species and $b \leq 1$ is a quasar radiation beaming
factor. (Strongly super-Eddington inflow may lead to significant beaming
\citep[e.g.][]{King2009MNRAS}, and we consider its effects in the
Discussion.) For the moment we keep $b$ as a free parameter. We assume that the
disk does not contribute to ionization; this is reasonable considering that
the disk radiates in the UV and longer wavelengths. Had we assumed the disc
spectrum to be the same as the SMBH spectrum, an extra factor $1 + {\rm
  ln}\left(1 + \dot{m}\right)$ would have appeared in the expression for
$\xi$.

Evaluating the constants in eq. (\ref{eq:xi}) gives
\begin{equation}
\xi_{\rm w} \simeq 5.7 \times 10^4 \eta_{\rm 0.1}^2 \frac{1 + {\rm
    ln}\left(1 + \dot{m}\right)}{\dot{m}^{2}} b \frac{l_{\rm
    i}}{10^{-2}},
\label{eq:xi2}
\end{equation}
where we have parametrized $\eta_{\rm 0.1} \equiv \eta / 0.1$. Crucially, none
of the wind parameters depend on the Eddington ratio of the AGN directly, but
only through $\dot m$. The high ionization parameter predicted for $\dot m
\sim 1$ explains why UFO winds are generally detected through FeXXV and
FeXXVI absorption lines \citep[cf.][]{King2010MNRASb}.

Equation (\ref{eq:xi2}) implies a set of self-consistent solutions for $\xi,
l_{\rm i}$ specified by $\dot m$, $b$ and $\eta$.  The ionization parameter
$\xi_{\rm w}$ depends linearly on the ionizing fraction of the AGN
luminosity. However, species of higher ionization level (corresponding to
higher ionization parameter) have higher threshold energies for ionization,
leading to a lower ionizing fraction $l_{\rm i}$. Hence, for a given set of
parameters $b$, $\eta$ and $\dot{m}$, there is at most one possible solution
where the ionization parameter calculated from eq. (\ref{eq:xi}) corresponds
to the ionized species as specified by $l_{\rm i}$. Conversely, different
combinations of these parameters can lead to different solutions of the
ionization equilibrium. In the next section, we estimate the ionization
equilibria for seven elements corresponding to various parameter ranges. We
will show that our model predicts the ionized species observed in BAL QSOs to
move with velocities consistent with observations.

\section{Model and results} \label{sec:model}

\subsection{Numerical model}

To find the connection between gas inflow parameters and self-consistent wind
ionization solutions, we use a simple numerical approach. We consider seven
elements commonly observed in BAL QSO spectra -- carbon, nitrogen, oxygen,
magnesium, silicon, sulphur and iron. Their ionized species cover a large
range of BAL QSO parameters.

We start by constructing a `typical' quasar SED, using data from
\citet{Elvis1994ApJS} and \citet{Winter2012ApJ}. The SED is composed of five
components: a rising radio and sub-mm continuum, flat IR, optical and UV
background with thermal IR and UV bumps superimposed, and a cut-off
power-law at X-ray and higher energies. We varied the SED to account for the
observational uncertainties and found that the overall results remain
unchanged. Using the SED, we find the fraction of total luminosity
that can ionize a particular element to any given level:
\begin{equation}
l_{\rm i,j} = \int_{E_{\rm i,j}}^{\infty} E_{\rm SED} {\rm d}E,
\end{equation}
where $E_{\rm SED}$ is the energy at a given energy and integration is carried
out above the ionization threshold of level i for element j. The SED is
normalized by construction.

Next, we consider the fractional abundances of various elemental ionization
levels in astrophysical plasmas as function of temperature. We use data from
\citet{Jordan1969MNRAS} for all elements except iron; for the latter, we take
the data from \citet{Arnaud1992ApJ}, since \citet{Jordan1969MNRAS} does not
give a complete ionization table for the element. We convert each temperature
to a corresponding ionization parameter using the prescription based on
\citet{Sazonov2005MNRAS}. The prescription is valid for an optically thin
plasma illuminated by, and in thermal equilibrium with, a quasar radiation
field; both assumptions are satisfied when considering equilibrium ionization
structures in a diffuse wind. In the few cases where temperatures fall outside
the range considered by \citet{Sazonov2005MNRAS}, we adopt a simplified
relation $\xi = T/200$ (in cgs units), which follows the approximate
analytical relation \citep{Sazonov2004MNRAS}. Given the fractional abundances
and the corresponding ionization parameters, we calculate the weighted average
ionization parameter for each ionization level of each element. This is the
$\xi_{\rm w}$ required for equilibrium.

We then substitute the calculated values of $\xi_{\rm w}$ and $l_{\rm i,j}$
into equation (\ref{eq:xi2}) and isolate the quantity
\begin{equation}
\frac{1 + {\rm ln}\left(1 +
  \dot{m}\right)}{\dot{m}^{2}} = \frac{\xi_{\rm w}}{5.7 \times 10^6
  l_{\rm i,j}b \eta_{\rm 0.1}^2}
\label{eq: xi3}
\end{equation}
for each element. We now use eqs. (\ref{eq:vwind}) and (\ref{eq: xi3})
to find $v_{\rm w}$ and $\dot m$ numerically for $b = 0.01, 0.1, 1$, assuming
$\eta = 0.1$.

\subsection{Extreme UFO solution}

\begin{figure}
  \centering
    \includegraphics[width=0.45 \textwidth]{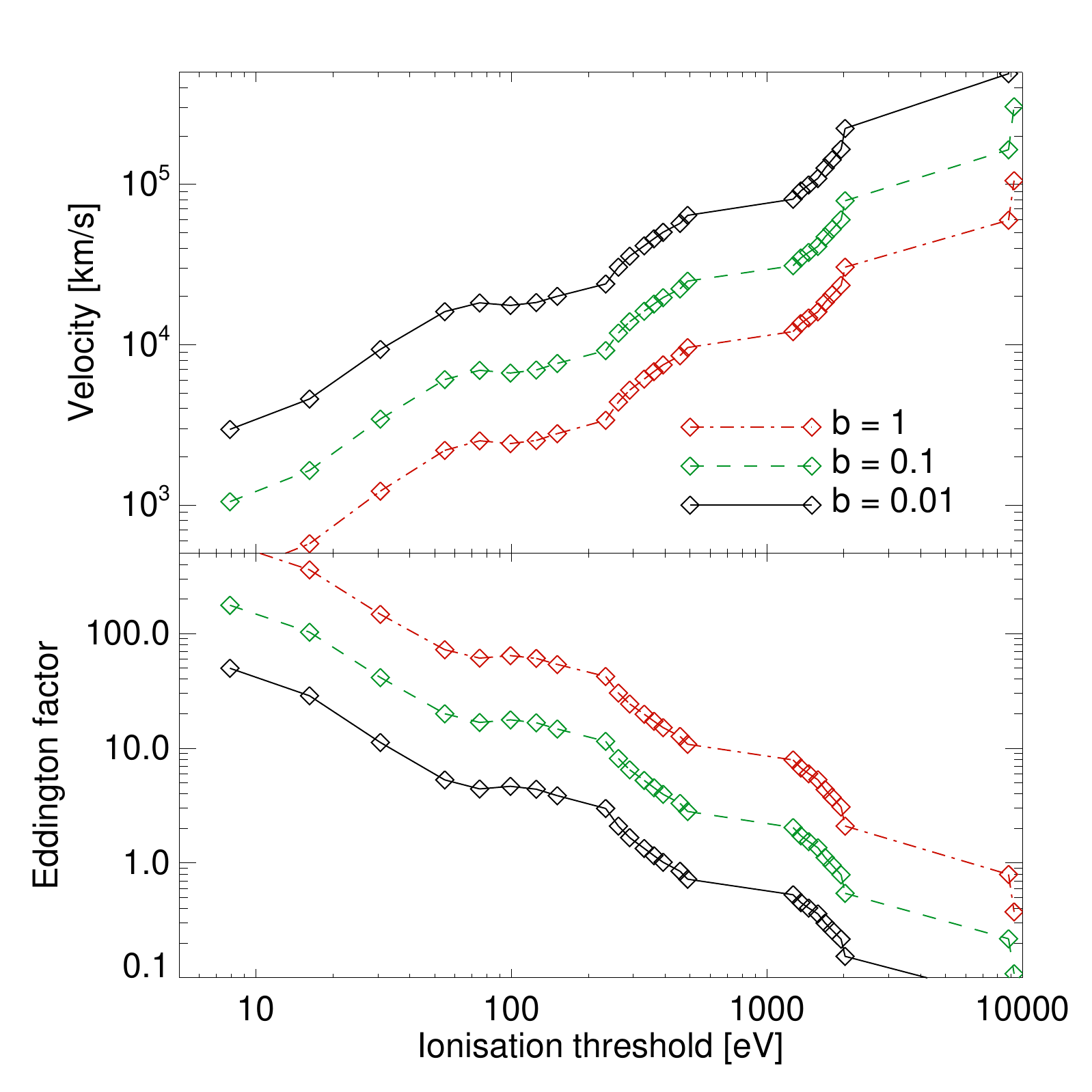}
  \caption{Wind velocity (top) and Eddington factor (bottom) as function of
    iron ionization level (diamonds) for three values of the luminosity
    beaming factor: $b = 0.01$ (black solid curve) $0.1$ (green
    dashed) and $1$ (red dot-dashed).}
  \label{fig:ionFe}
\end{figure}

Figure \ref{fig:ionFe} shows the outflow velocity and Eddington factor
given by eqs. (\ref{eq:vwind}) and (\ref{eq: xi3}) for all iron
species. Each diamond corresponds to a different ionization level,
increasing from FeII to FeXXVI. The three curves are for $b=0.01$
(black solid), $b=0.1$ (green dashed) and $b=1$ (red
dot-dashed). 

The two rightmost points in the figure are for helium- and
hydrogen-like iron ions, and describe mildly relativistic extreme UFO
winds. For the expected unbeamed luminosities ($b = 1$) the outflow
has $\dot{m} \lesssim 1$, i.e. the accretion rate is not much higher
than Eddington. The predicted outflow velocities are $\sim 0.15 -
0.3c$, within the observed range \citep{Tombesi2010A&A,
  Tombesi2010ApJ}.

\subsection{BAL QSO solution}

\begin{figure}
  \centering
    \includegraphics[width=0.45 \textwidth]{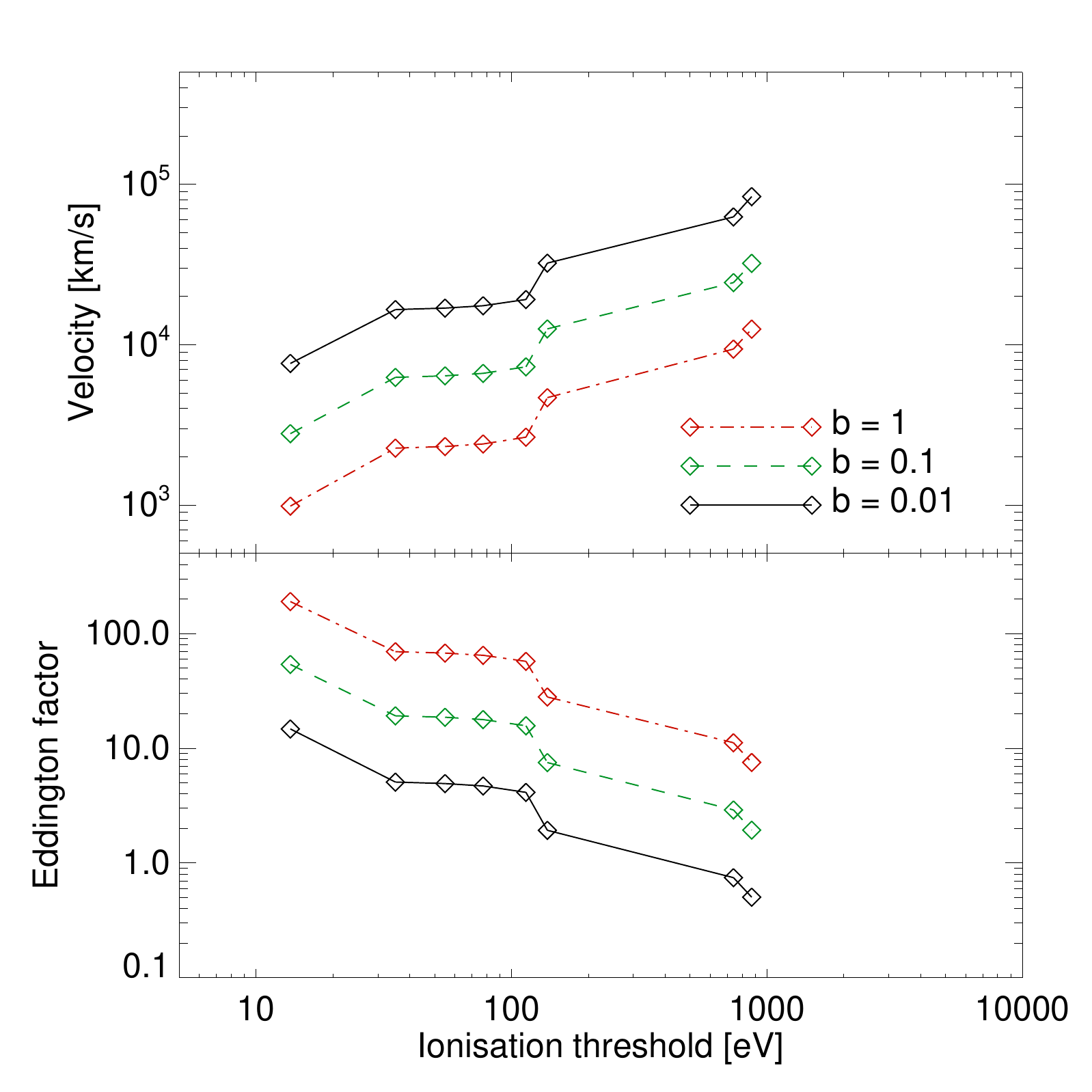}
  \caption{Same as Figure \ref{fig:ionFe}, but for oxygen.}
  \label{fig:ionO}
\end{figure}

In Figure \ref{fig:ionO}, we plot the same data as in Figure \ref{fig:ionFe},
but for oxygen ionization states. BAL QSOs are often observed via an OVI
absorption line \citep[e.g.][]{Baldwin1996ApJ}, so we concentrate on the
results for this ion (5th diamond from the left). It is clear that the major
difference from Seyfert winds is the significantly higher Eddington factor
$\dot m \sim 10$. For flows with spherically symmetric radiation fields ($b =
1$), this ionization state is consistent with $v_{\rm w,O} \simeq 2500~{\rm
  km\, s}^{-1}$, close to the lower bound of the minimum BAL QSO outflow
velocities \citep{Gibson2009ApJ}. However, if the Seyfert luminosity is beamed
perpendicular to the disk plane, the velocity is higher. For a beaming factor
$b = 0.1$, we find $v_{\rm w,O} \simeq 7000$~${\rm km\,s}^{-1}$, well within
the range of observed BAL QSO parameters. Other ions -- CIV, NV, SiIV, SIV and
SVI -- show very similar results. We compare our results with the range of
minimum, rather than the maximum, BAL QSO outflow velocities, because we only
consider radiative acceleration of winds. The wind is usually launched with
considerable velocity (similar to the escape velocity at the launch radius) by
gas pressure, which would increase its final velocity as well.

\begin{figure}
  \centering
    \includegraphics[width=0.45 \textwidth]{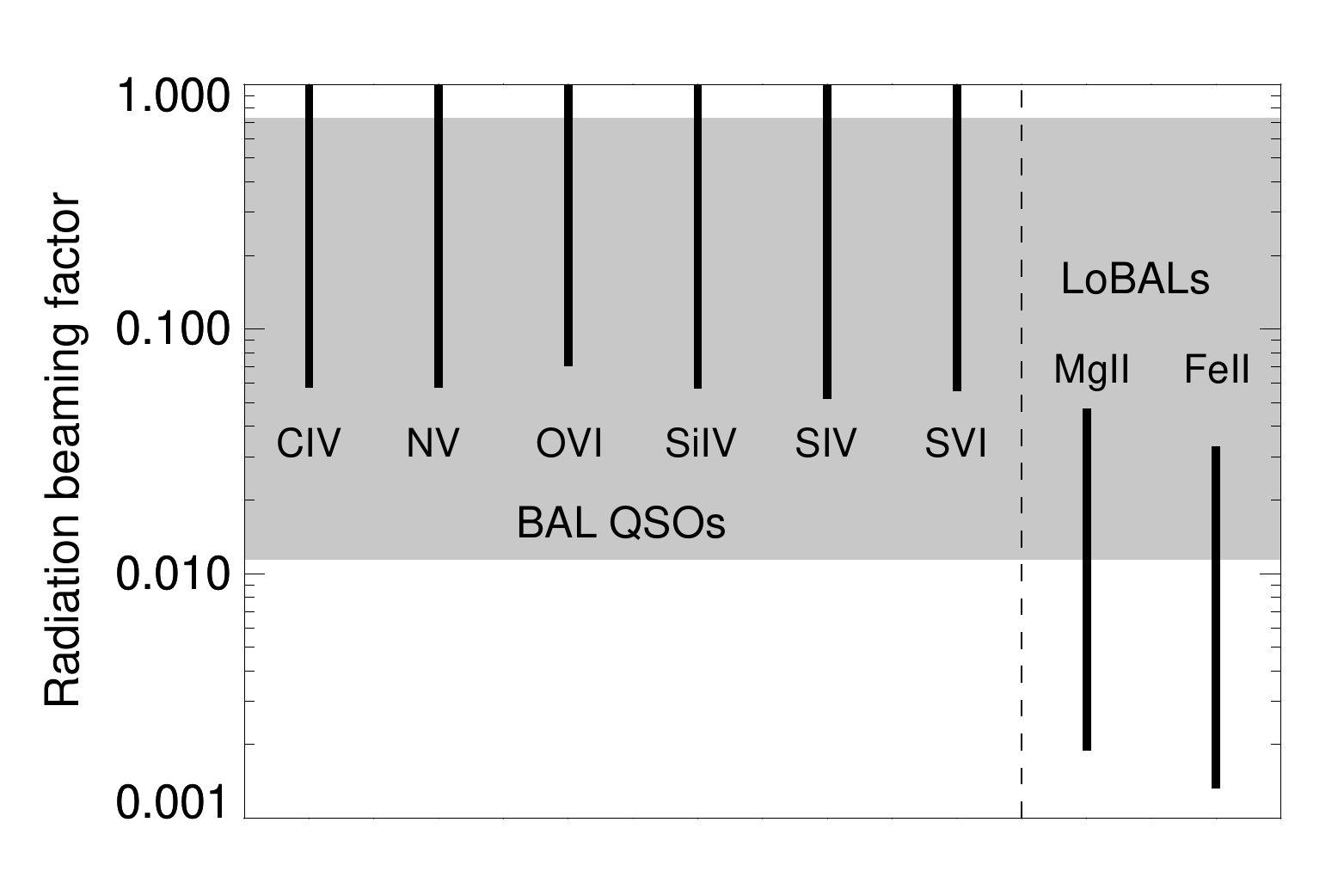}
  \caption{Range of beaming factors giving ionization equilibrium solutions
    consistent with the lower end of BAL QSO wind velocities $v_{\rm w}
    = 2000-10^4~{\rm km\,s}^{-1}$ for commonly observed BAL QSO ions. Higher
    beaming factors correspond to lower velocities. Low ionization species
    MgII and FeII are seen in a small fraction of BAL QSOs, suggesting a
    short-lived evolutionary phase.}
  \label{fig:branges}
\end{figure}

To quantify the constraints on luminosity beaming for BAL QSO
outflows, we plot $b$ for eight common ions giving wind velocities
$v_{\rm w} = 2000-10^4$~${\rm km\,s}^{-1}$ (Figure
\ref{fig:branges}). The six ions most commonly observed in BAL QSOs --
CIV, NV, OVI, SiIV, SIV and SVI -- all have solutions with $0.08
\lesssim b \lesssim 1$. The shaded region corresponds to the range of
beaming factors predicted by equation (8) of \citet{King2009MNRAS} for
$\dot m$ values between $10$ and $80$.

\subsection{LoBAL solution}

The presence of two more ions -- MgII and FeII -- in this simple picture is
consistent with the observed outflow velocities only if the AGN radiation
field is strongly beamed, with $b \simeq 0.01$. Such strong beaming may not be
required if other processes can reduce the ionization level in these systems,
but our simple model does not take such effects into account. We comment
further on the validity of this model and several possible complicating issues
in the next Section.

\section{Discussion} \label{sec:discuss}

\begin{sidewaysfigure*}
%\begin{figure*}
  \centering
  \vspace{8cm}
    \includegraphics{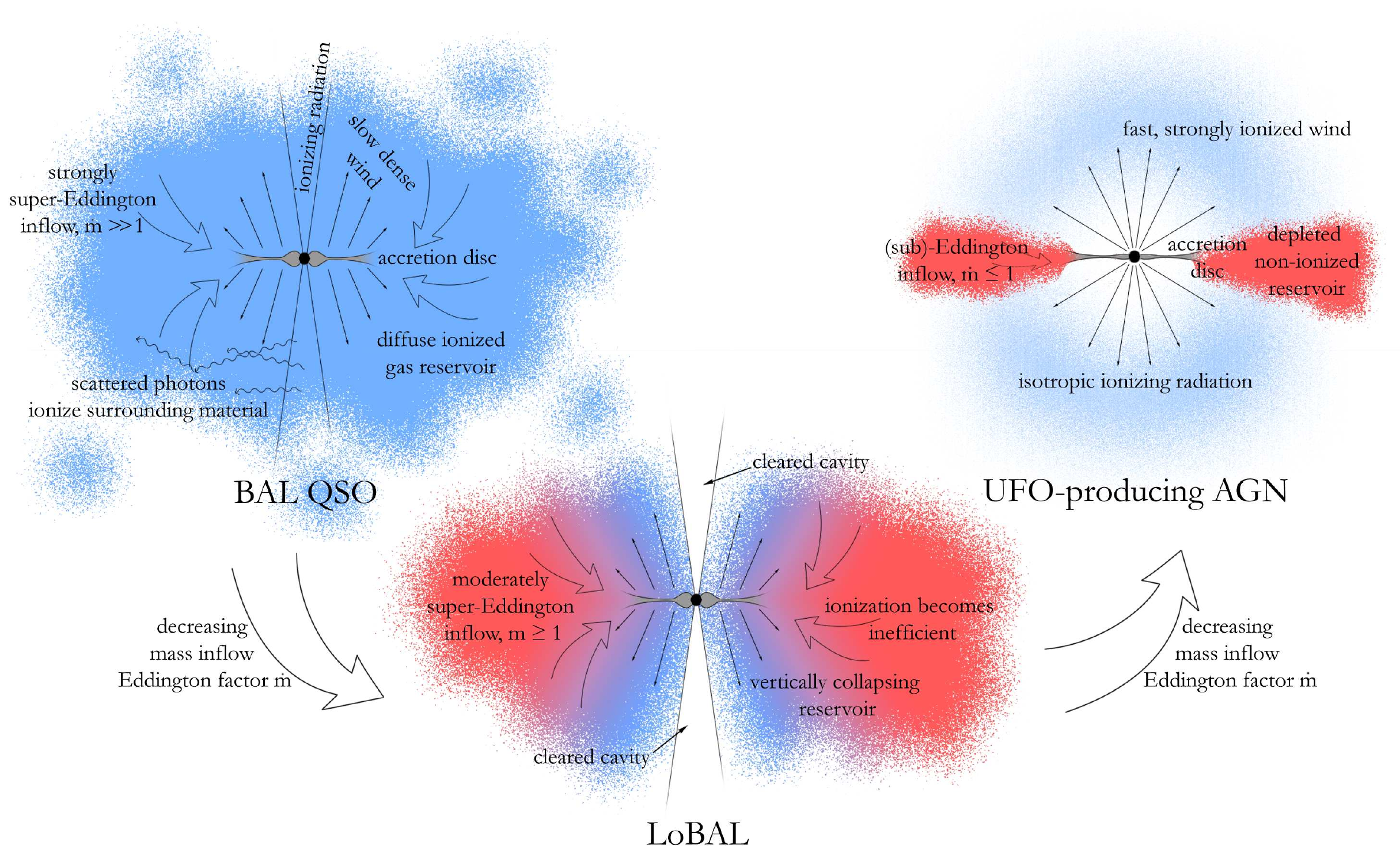}
  \caption{Suggested schematic evolution of a bright accretion event. BAL QSOs
    correspond to the initial state of highly super-Eddington accretion and
    outflow (left panel). The gas reservoir is ionized (shown in blue) by
    scattered radiation from the quasar. In LoBALs, the ionizing radiation has
    cleared a cone in the gas reservoir so that most of the wind gas sees few
    ionizing photons (non-ionized gas shown in red). In UFO-producing AGN, the
    gas reservoir is strongly depleted and the infall rate close to {\bf or
      below} Eddington, allowing high ionization of the dilute outflowing
    wind.}
  \label{fig:cartoon}
%\end{figure*}
\end{sidewaysfigure*}

The results of the last Section suggest that the main parameter
affecting the properties of an outflowing radiation--pressure driven
wind in AGN is its Eddington factor $\dot m$, and so ultimately the
inflow rate from the host galaxy towards the central supermassive
black hole. This is $\sim 1$ in extreme UFO-producing AGN, but $\simgt
10$ in BAL QSOs. We can draw several conclusions from this.

\subsection{BAL QSO fraction}

BAL QSOs comprise a small fraction, $\sim 15 \%$, of all AGN
\citep{Knigge2008MNRAS}. In our picture this follows from the requirement for
inflow from the host galaxy at significantly super--Eddington rates. From
eq. (\ref{edd}) these conditions are more easily achieved in galaxies with low
values of $\sigma$ or SMBHs with masses well below the $M-\sigma$ relation,
but both cases correspond to low absolute luminosities and so are disfavoured
by observation. So the rarity of BAL QSOs suggests that near--dynamical inflow
rates in big galaxies with big black holes are rare, presumably because such
episodes are shortlived.

\subsection{Absorption line widths}

The fact that the accretion disk contributes significantly to driving the wind
may explain the width of BAL QSO absorption lines. Since the wind is
launched from a wide range of radii in the disc, its gas has a wide range of
initial velocities. Even though they are all accelerated by the central
source, the original spread remains.

\subsection{Warm absorbers}

A large fraction of Seyfert galaxies, as well as some quasars, show absorption
in the UV and soft X-ray spectra of material moving with comparatively low
velocities, $v_{\rm WA} = 10-1000$~km/s \citep{McKernan2007MNRAS} with line
widths similar to these velocities. They are composed of weakly ionised ($\xi
\simeq 30$~erg cm s$^{-1}$) $10^5$~K gas
\citep[e.g.,]{Reynolds1995MNRAS}. Ionization models show that these ``warm
absorbers'' exist outside the broad line regions of their AGN and are perhaps
associated with the gas in the torus \citep{Blustin2005A&A}. Comparison of
these systems with typical Seyferts suggests that they should be accreting
with high Eddington ratios $\dot m$ \citep{Brandt2000ApJ}.

The model presented in this paper does not account for multiple ionization
levels in an outflow. Nevertheless, a simple qualitative picture can be
developed to explain these. The wind rising from the accretion disc around an
AGN is stratified in density, with slower and denser material close to the
disc and faster lower density material further away. As this wind is
illuminated by an isotropic AGN radiation field, its ionization parameter
increases with height above the disc mid--plane. For a Gaussian vertical
density profile, the particle density drops (and the ionization level
increases) by a factor $1000$ within less than 3 scale heights from the base
of the wind; this difference would allow both warm absorbers and hydrogen-like
iron to coexist within the same outflow. A similar model was proposed for the
weak outflow in NGC 5548 \citep{Steenbrugge2005A&A}; we claim that this is the
general case.

In addition, any inhomogeneities present in the ISM surrounding the AGN can
produce lower ionization features in the spectrum. Such features of varying
velocities and ionization levels have been identified in both UFOs
\citep[e.g.,]{Pounds2011MNRAS} and BAL QSOs \citep[e.g.,]{Moe2009ApJ}.
Finally, it is possible that low--velocity, low--ionization species are formed
by recombination in the coolest part of the postshock flow (Pounds \& King,
2013, in preparation).

\subsection{What are LoBALS?}

Our work suggests that AGN winds and BAL QSO outflows are explicable by
the same physical mechanism, the only difference being the
large-scale properties of the reservoir feeding the black hole. The
LoBALs appear to be a rare state in which the AGN radiation is somehow
beamed away from most of the outflow.

We speculate that AGN may evolve over time from an initial BAL QSO state of an
AGN, through a LoBAL state, to a final stage as a Seyfert galaxy or a quasar
(see Figure \ref{fig:cartoon} for a schematic of the process). We assume that
an AGN accretion episode starts when some process creates a large reservoir of
gas of relatively low angular momentum around the SMBH (left panel of the
Figure). This would give a near-dynamical inflow rate $\dot M_{\rm in} \la
M_{\rm dyn}$ and hence a highly super-Eddington SMBH accretion rate ($\dot{m}
\gg 1$). The black hole and its accretion disk begin driving an outflow, which
is observed in broad absorption lines, with the accretion luminosity being
somewhat beamed ($b \ga 0.1$), giving a BAL QSO.

The radiation pressure inside the beamed ionizing cone is $p_{\rm
  r,beam} \propto b^{-1}$, where $b$ probably decreases with $\dot m$
(\citealt{King2009MNRAS} suggests that $ b \propto \dot{m}^{-2}$),
reinforcing the tendency of larger accretion rates to give higher
radiation pressure.  The pressure inside the ionization cone is much
higher than the pressure in the more modestly illuminated vicinity of
the disc, $p_{\rm r,d} \propto \left(1 + {\rm ln} \left(1 +
\dot{m}\right)\right)$. As a result, the beamed ionizing radiation
flux evacuates a conical cavity from the surrounding material. This
process, together with the inevitable vertical collapse of the
large-scale reservoir as it circularizes close to the SMBH, reduces
the effective beaming factor severely (to $b \sim 10^{-2}$) as there
is less material interacting with the beamed radiation of the
SMBH. This stage corresponds to a LoBAL (middle panel).

The large-scale reservoir feeding the AGN is depleted in a few dynamical
times $t_{\rm d} \simeq 5 \times 10^4 R_{10} \sigma_{200}^{-1}$~yr, with
$R_{10}$ the size of the reservoir in tens of parsecs. As this happens, the
Eddington factor $\dot{m}$ drops until eventually only the accretion disk
remains. The SMBH then starts to accrete at $\dot{m} \simeq 1$, producing a
UFO--type wind.

Both changes to the system state depend on dynamical processes (reservoir
collapse and depletion), so they happen soon after one another. This may
explain why LoBALs are rare -- they are systems in a process of changing from
BAL QSOs to more modestly accreting AGN.

\subsection{A particular example: SDSS J1106+1939}

A recently discovered powerful BAL QSO outflow in SDSS J1106+1939
\citep{Borguet2012arXiv} has a mass flow rate of $400 \; \msun$~yr$^{-1}$ and
velocity $v \simeq 8000$~${\rm km\,s}^{-1}$. From equation (\ref{eq:vwind}) we
find that this velocity corresponds to $\dot{m} = 14$, while
eq. (\ref{eq:lkin}) gives $L_{\rm kin} = 0.05 L_{\rm Edd}$ for this system,
exactly as observed. The Eddington ratio also implies an Eddington accretion
rate of $\sim 29 \; \msun$~yr$^{-1}$, giving an SMBH mass $M_{\rm
  SMBH} = 1.3 \times 10^9 \; \msun$, almost exactly equal to the $\sim 1.5
\times 10^9 \; \msun$ SMBH mass calculated from the observed kinetic
luminosity. 

\citet{Borguet2012arXiv} determine the scale of the flow as about 300~pc,
considerably smaller than the likely Compton cooling radius \citep[$\sim
  3$~kpc for a black hole of mass $M_{\rm SMBH} = 1.3 \times 10^9 \; \msun$
  lying close to the $M - \sigma$ relation; see][]{Zubovas2012MNRASb}. So it
is likely that the collision of the observed wind with the host gas will
result in a strongly cooled shock, and not sweep the galaxy clear of gas in an
energy-driven, possibly molecular, outflow.

\section{Conclusion}

We have shown that the outflowing winds in Seyfert galaxies and BAL QSOs
differ only in their mass outflow rates. These are of order the
Eddington accretion rate in Seyferts, but $\sim 10-50$ times this
in BAL QSOs. Our picture correctly predicts the velocities and
ionization states of the observed winds, including the
recently-discovered case of SDSS J1106+1939. We suggest that luminous
AGN may evolve from BAL QSO through LoBAL to Seyfert as their
Eddington factors drop during the decay of a bright accretion
event. LoBALs correspond to a short-lived stage in which the AGN
radiation pressure largely evacuates the ionization cone, but before
the large-scale accretion rate has dropped to Eddington value.

We note finally that equations (\ref{eq:ldisc},\ref{eq:eddthrust})
imply that extreme UFOs and BAL QSO winds exert the Eddington thrust
on the interstellar gas of their host galaxies, and so produce the $M
- \sigma$ relation (\ref{eq:msigma}), which agrees with
observation. The analogous equations
(\ref{eq:ldisc},\ref{eq:subeddthrust}) for {\it sub}--Eddington UFOs
would instead produce an $M - \sigma$ relation with the black hole
masses {\it larger} by factors $\simeq \le/L_{\rm bol}$, which can
approach 100 in some cases. Since larger masses should be easier to
measure than smaller ones, it seems unlikely that such a relation
holds in reality. We therefore conclude that such sub--Eddington
systems cannot be the sites of significant SMBH mass growth. Unless
this occurs in obscurity, most galaxies must pass through prolonged
phases as extreme UFOs or BAL QSOs.

\section*{Acknowledgments}
 
We thank Mike Goad and Belinda Wilkes for enlightening discussions
regarding observed features of BAL QSOs. We are particularly grateful
to the anonymous referee for bringing to our attention several issues
that helped us improve the paper considerably. KZ was supported by an
STFC studentship and later by an STFC Rolling Grant for Theoretical
Astrophysics.

\end{document}